\documentclass[twocolumn,prl]{revtex4-1}
\usepackage[latin9]{inputenc}
\usepackage{xcolor}
\usepackage{array}
\usepackage{textcomp}
\usepackage{amsmath}
\usepackage{amssymb}
\usepackage{graphicx}
\usepackage[unicode=true,pdfusetitle,
 bookmarks=true,bookmarksnumbered=false,bookmarksopen=false,
 breaklinks=false,pdfborder={0 0 0},pdfborderstyle={},backref=false,colorlinks=true]
 {hyperref}
\hypersetup{
 colorlinks,linkcolor=red,citecolor=blue}

\makeatletter

\newcommand*\LyXThinSpace{\,\hspace{0pt}}
\DeclareFontEncoding{LGR}{}{}

\ProvideTextCommand{\~}{LGR}[1]{\char126#1}

\DeclareTextSymbolDefault{\textquotedbl}{T1}
\providecommand{\tabularnewline}{\\}

\usepackage{amsfonts}

\makeatother

\begin{document}
\title{Photonic chip-based high-efficiency soliton microcombs via electroopitc-Kerr
synergy}
\author{Rui Niu$^{1,2,5,6,7,\dagger}$, Shuai Wan$^{1,2,5,6,\dagger}$, Pi-Yu
Wang$^{1,2,5,6}$, Rui Ma$^{3}$, Jin Li$^{1,2,5,6}$, Fang Bo$^{3,*}$,
Zhen Shen$^{1,2,5,6}$, Guang-Can Guo$^{1,2,5,6}$, Fang-Wen Sun$^{1,2,5,6,*}$,
Junqiu Liu$^{4,6,*}$, Chun-Hua Dong$^{1,2,5,6,*}$}
\affiliation{$^{1}$Laboratory of Quantum Information, University of Science and
Technology of China, Hefei, Anhui 230026, China}
\affiliation{$^{2}$CAS Center For Excellence in Quantum Information and Quantum
Physics, University of Science and Technology of China, Hefei, Anhui
230088, China}
\affiliation{$^{3}$MOE Key Laboratory of Weak-Light Nonlinear Photonics, TEDA
Applied Physics Institute and School of Physics, Nankai University,
Tianjin 300457, China}
\affiliation{$^{4}$International Quantum Academy, Shenzhen 518048, China}
\affiliation{$^{5}$Anhui Province Key Laboratory of Quantum Network, University
of Science and Technology of China, Hefei, Anhui 230088, China}
\affiliation{$^{6}$Hefei National Laboratory, University of Science and Technology
of China, Hefei, Anhui 230088, China}
\affiliation{$^{7}$Present address: School of Physics, Harbin Institute of Technology,
Harbin 150006, China}
\affiliation{$^{\dagger}$These authors contributed equally to this work.}
\email{bofang@nankai.edu.cn, fwsun@ustc.edu.cn, liujq@iqasz.cn, chunhua@ustc.edu.cn}

\begin{abstract}
Temporal soliton mode-locking in coherently pumped microcavities provides
a promising platform for miniaturized frequency comb systems. While
significant progress has been made, achieving high conversion efficiency
in such microcombs remains a critical challenge. Soliton generation
through pulse pumping has emerged as an effective strategy to improve
conversion efficiency. However, the on-chip integration of pulse generation
with dissipative Kerr soliton (DKS) formation within the photonic
chip has not yet been realized. In this work, we demonstrate a photonic
chip-based soliton microcomb with high conversion efficiency, achieved
by integrating on-chip pulse generation and DKS generation. The pulsed
laser, fabricated on a lithium niobate-on-insulator (LNOI) platform,
delivers a $35.5\LyXThinSpace\mathrm{GHz}$ repetition rate with broadly
tunable center frequencies. By coupling these on-chip pulses to a
silicon nitride microresonator, we achieve stable DKS generation with
a pump-to-soliton conversion efficiency of $43.9\,\text{\%}$ under
steady-state conditions. This integrated architecture establishes
a viable pathway toward chip-scale soliton microcombs with unprecedented\textbf{
}efficiency, opening up new possibilities for optical communications,
precision spectroscopy, and photonic sensing.
\end{abstract}
\maketitle

\section{INTRODUCTION}

The rapid advancement of integrated photonics has notably energized
research into the generation of dissipative Kerr solitons (DKSs) on
photonic chips, leveraging third-order nonlinearity within microresonators
\cite{1kippenberg2018dissipative,wang2020advances}. DKSs offer a
promising route to achieve broadband, fully coherent, chip-scale frequency
combs with advantages of compact, lightweight, and energy-efficient
\cite{shen2020integrated,chang2022integrated}. Unlike conventional
mode-locked lasers, soliton microcombs operate over a wide range of
repetition rates, from the microwave to the millimeter-wave domains
\cite{liu2020photonic,tetsumoto2021optically}. Various material platforms
such as silica \cite{bai2021brillouin,yao2022soliton,niu2023khz},
silicon nitride \cite{rao2021towards,moille2023kerr}, silicon \cite{yu2018silicon},
lithium niobate \cite{he2023high,yang20231550,wan2024photorefraction,lv2025broadband},
aluminum nitride \cite{bruch2021pockels}, and silicon carbide \cite{wang2022soliton,guidry2022quantum}
have hosted demonstrations of DKS generation. The integration of photonics
has played a pivotal role in enhancing the capabilities of soliton
microcomb technology, significantly accelerating its adoption across
various system-level applications such as telecommunications , ultrafast
ranging \cite{chen2023breaking,shi2024frequency,zhu2025integrated},
microwave generation \cite{sun2024integrated,niu2024ultralow,zhao2024all},
dual-comb spectroscopy \cite{bao2021architecture,li2024coherently},
quantum information \cite{huang2025massively}, frequency synthesizers
\cite{spencer2018optical}, and optical atomic clocks \cite{wu2025vernier}.

\begin{figure*}[t]
\centering{}\includegraphics[clip,width=1.8\columnwidth]{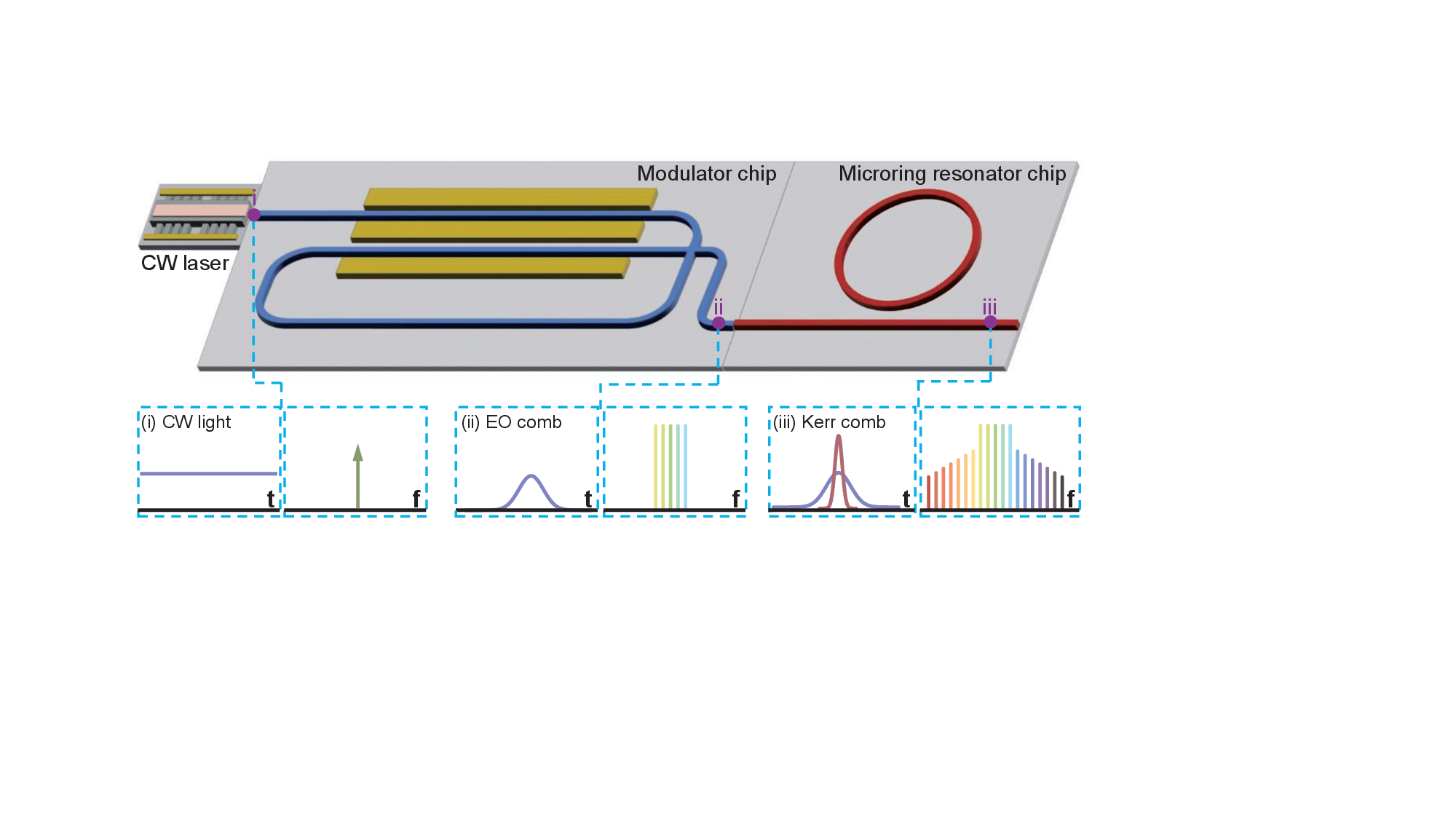}\caption{\label{fig:1} The schematic of the photonic chip-based high conversion
efficiency soliton microcomb. A continuous-wave is coupled into the
lithium niobate photonic chip as a phase modulator for the generation
of electro-optic frequency comb, which is sent to an optical microcavity
to generate a DKS. The repetition rate of this electro-optic comb
closely matches the FSR of the optical microcavity. Insets (i-iii)
show the time domain and frequency domain of the CW light, EO comb
and Kerr comb.}
\end{figure*}

Despite their broad applicability, DKSs face several key challenges.\textbf{
}The\textbf{ }pump-to-soliton conversion efficiency\textbf{ }remains
relatively low, resulting in low pump energy utilization and high
pump power requirements \cite{helgason2021dissipative,li2022efficiency}.
Strong thermal effects in microresonators further complicate DKS generation
\cite{zhang2019sub}. Moreover, stabilizing the repetition rate of
DKSs requires fast photodetection and microwave signal processing,
making the system more complex \cite{newman2019architecture,drake2019terahertz}.

Pulse pumping of microcavities offers a straightforward solution to
these challenges. By enhancing the overlap between the pump pulse
and the soliton pulse, this approach improves conversion efficiency
\cite{obrzud2017temporal,zhang2025impact}. For instance, a silica
microcavity pumped with pulses achieved a $34\,\text{\%}$ conversion
efficiency at the steady-state \cite{li2022efficiency}. Similarly,
a silicon nitride-based microcomb with a detectable repetition rate
of $28\,\mathrm{GHz}$ demonstrated an $8\,\text{\%}$ conversion
efficiency and a 2/3-octave spectral span \cite{anderson2021photonic}.
Additionally, pulse pumping mitigates thermal effects, enabling DKS
generation through simple pump wavelength sweeping \cite{li2022efficiency}.
The repetition rate of the generated soliton is inherently locked
to that of the pump pulse, thereby simplifying system design \cite{obrzud2017temporal}.
However, in the aforementioned works, the driving pulses are generated
by bulky cascaded fiber-optic electro-optic modulators. 

Achieving on-chip pulse-driven DKSs, where both pulse generation and
DKS formation occur on the photonic chip, remains a significant challenge.
A critical requirement is the precise matching of the repetition rate
and center frequency of the on-chip pulse with those of the DKS. Electro-optic
(EO) modulation is a well-established technique for generating optical
pulses from CW lasers with tunable frequency and repetition rates
\cite{beha2017electronic,qi2020integrated,zhuang2023electro,niu2023integrated}.
Recent advancements in the lithium niobate on insulator (LNOI) platform
have enabled the generation of ultrashort on-chip pulses \cite{yu2022integrated,zhang2023power,guo2023ultrafast}.
While the frequency comb generation using a $\chi^{(3)}$ resonator
synchronously pumped by a tunable femtosecond pulse generator with
on-chip amplitude and phase modulators has been proposed \cite{cheng2024frequency},
the realization of combined on-chip pulse and DKS generation remains
elusive.

In this paper, we present a photonic chip-based soliton microcomb
with high conversion efficiency, achieved by integrating on-chip pulse
generation with DKS formation. The on-chip pulses are generated on
the LNOI platform with a $35.5\,\mathrm{GHz}$ repetition rate and
a tunable center frequency, while the DKSs are generated on a silicon
nitride ($\mathrm{Si_{3}N_{4}}$) platform with the approximate repetition
rate. Using the on-chip pulse to drive DKS generation minimizes thermal
effects in the microresonator and allows for soliton formation through
simple pump frequency scanning, achieving a steady-state single-soliton
with conversion efficiency of $43.9\,\%$. Our work demonstrates a
promising pathway toward high-efficiency, chip-scale soliton microcombs,
with potential applications in optical data transmission, spectroscopy,
and sensing.

\section{Schematic and device fabrication}

\begin{figure*}[t]
\centering{}\includegraphics[clip,width=2\columnwidth]{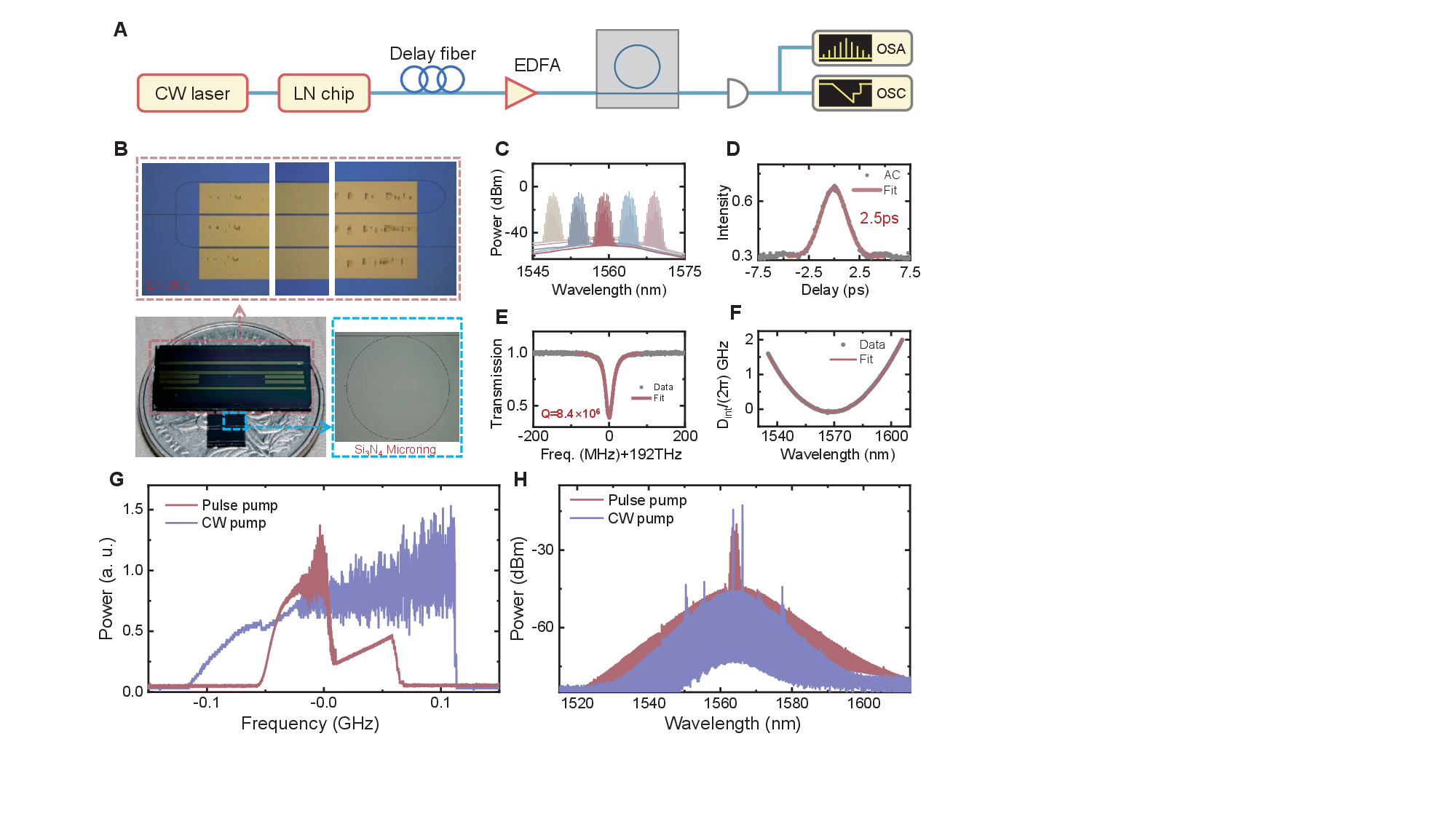}\caption{\label{fig:2} \textbf{(A)} Schematic of the experimental setup. EDFA:
erbium-doped fiber amplifier; OSC: oscilloscope; OSA: optical spectrum
analyzer.\textbf{ (B)} Photograph of the LN chip and $\mathrm{Si_{3}N_{4}}$
photonic chip. The microscope image of the device consisting of a
LN phase modulator with electrodes and $\mathrm{Si_{3}N_{4}}$ microring
resonator. \textbf{(C)} Optical spectrum of the EO comb before amplification
with various center wavelength. \textbf{(D}) Autocorrelation measurement
of the EO comb. \textbf{(E)} Resonance linewidth fitting of $\mathrm{Si_{3}N_{4}}$
microring resonator, showing loaded optical Q-factor of 8.4 million.
\textbf{(F)} The measured dispersion of the $\mathrm{Si_{3}N_{4}}$
microring resonator. \textbf{(G)} Measured soliton step when the CW
or pulse pump is scanning through the optical mode. \textbf{(H)} Optical
spectrum of the generated DKS with the CW and pulse pump. An auxiliary
laser around $1566\,\mathrm{nm}$ is used to compress the thermal
effect with the CW pump laser.}
\end{figure*}

The architecture of the photonic chip-based high-efficiency soliton
microcomb system is illustrated in Fig. \ref{fig:1}. This system
consists of a CW laser, an on-chip phase modulator, and a microring
resonator satisfying the conditions for soliton generation. The CW
laser (i) is coupled into the modulator chip, where cascaded EO modulation
generates an EO frequency comb. Simultaneously, the CW laser is modulated
in the time domain to form a pulsed laser (ii). The EO comb is then
coupled into the microring resonator, where the center frequency and
repetition rate of the comb are tuned to match the resonance modes,
thereby generating a pulse-pumped soliton microcomb (iii). The insets
depict the temporal and spectral characteristics at each stage of
the system. Compared to conventional CW pumping schemes, the pulse-pumped
approach provides higher peak power and better temporal overlap with
soliton pulses. Consequently, the pump-to-comb conversion efficiency
is significantly enhanced. The maximum conversion efficiency can be
expressed as \cite{li2022efficiency}:

\begin{equation}
\varGamma_{\mathrm{pulse}}=2\pi^{2}\eta^{2}\tau/\tau_{\mathrm{p}}\label{eq:eq1}
\end{equation}

where $\tau$ and $\tau_{\mathrm{p}}$ represent the duration of the
soliton and the pump pulse, respectively, and $\eta=Q/Q_{\mathrm{ex}}$
represents the resonator loading factor, with $Q$ and $Q_{\mathrm{ex}}$
being the total and external quality factors. According to this formula,
improving the conversion efficiency requires not only maximizing the
Q-factor of the microring resonator but also narrowing the pump pulse
width (i.e., broadening the spectrum of the electro-optic comb).

Here, we utilized the on-chip phase modulator based on the LNOI platform
to generate EO comb. For the pump temporal profile assumed here \cite{li2022efficiency},
we also have $\tau_{\mathrm{p}}=T_{\mathrm{R}}/N$, where N is the
number of spectral lines in the pump, $T_{R}$ is the cavity round-trip
time, and where pumping close to the round-trip rate of the microcomb
is assumed. In this case, the maximum efficiency can be estimated
on the form

\begin{equation}
\varGamma_{\mathrm{pulse}}=N\varGamma_{\mathrm{CW}}\label{eq:eq2}
\end{equation}
where $\varGamma_{\mathrm{CW}}$ is the maximum cw pumping efficiency.
According to this formula, the efficiency of the pulse pumping mainly
dependeds on the comb-line number of the EO comb. Furthermore, we
generate the DKS in the microring resonator based on the $\mathrm{Si_{3}N_{4}}$
platform, which offers higher Q-factors while ensuring reliable soliton
microcomb generation.

The on-chip phase modulator was fabricated using a commercial x-cut
LNOI wafer provided by NANOLN. This wafer consists of a 600 nm-thick
x-cut LN layer, a 2 \textmu m-thick wet-oxidized silicon dioxide ($\mathrm{SiO{{}_2}}$)
layer, and a 500 \textmu m-thick silicon substrate. Detailed fabrication
processes can be found in \cite{wan2024photorefraction}. After preparing
the LN waveguides, microwave electrodes were fabricated via laser
direct writing, followed by metal deposition using thermal evaporation.
A dual-layer lift-off process was employed to transfer a $15\,\text{nm}$
chromium layer and a $300\mathrm{\,nm}$ gold layer. The microwave
electrodes are $1.9\,\mathrm{cm}$ long, and the LN waveguide passes
through the electrodes twice to enhance the modulation efficiency.

\begin{figure*}[t]
\begin{centering}
\includegraphics[clip,width=1.5\columnwidth]{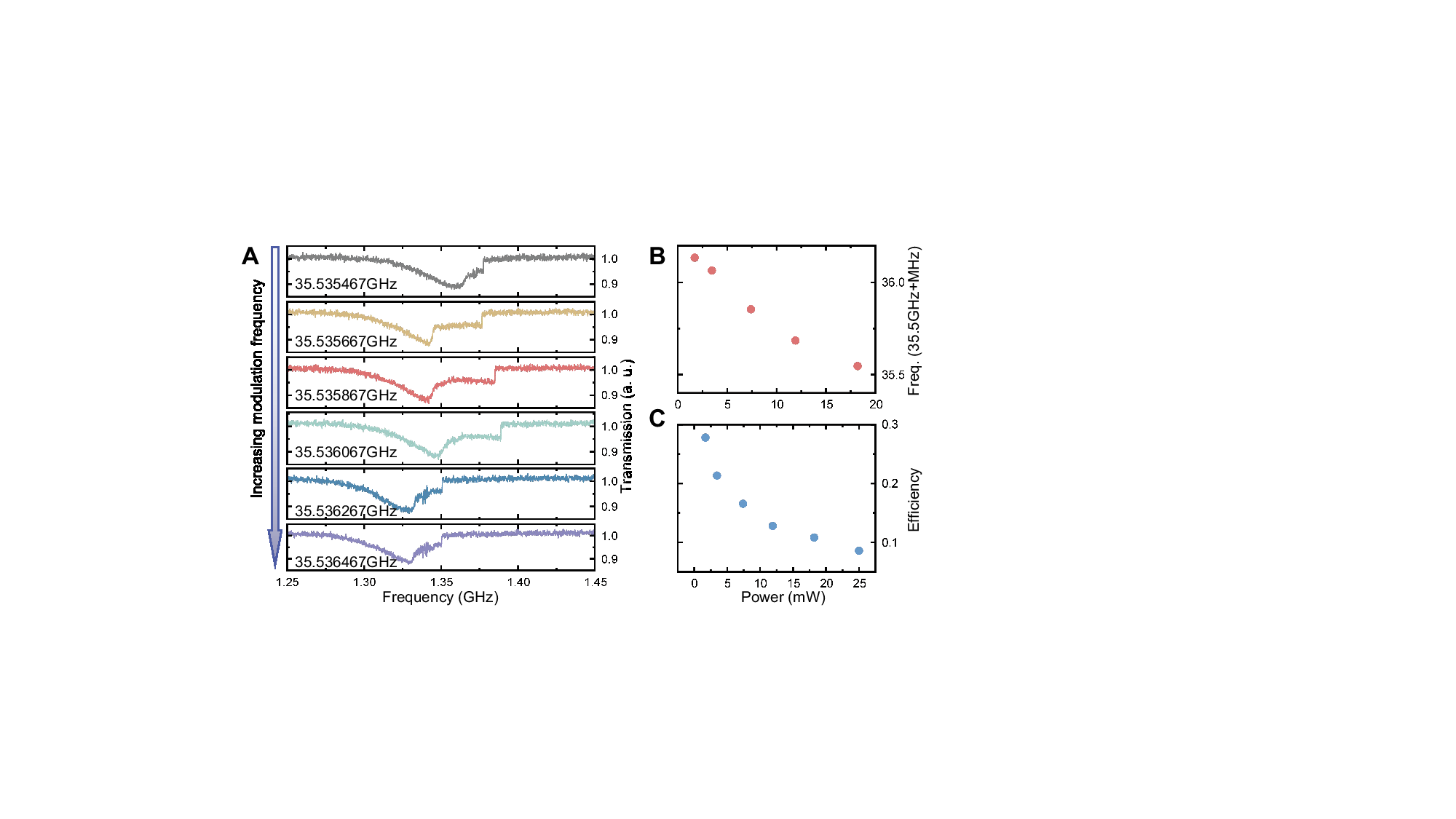}
\par\end{centering}
\centering{}\caption{\label{fig:3} \textbf{(A) }The transmission of the pump laser as
the modulation frequency of the on-chip pulse increases from $35.535467\,\mathrm{GHz}$
to $35.536467\,\mathrm{GHz}$. \textbf{(B)} Measured repetition rate
versus the pump power, with the increasing of the pump power, the
repetition rate decreases.\textbf{ (C)} Measured conversion efficiency
versus the pump power, with the increasing of the pump power, the
conversion efficiency decreases.}
\end{figure*}

The $\mathrm{Si_{3}N_{4}}$ microresonator was fabricated using a
subtractive process \cite{ye2023foundry} with 830-nm-thick $\mathrm{Si_{3}N_{4}}$
on 150-mm-diameter (6-inch) wafers. The process started with deposition
of 830-nm-thick $\mathrm{Si_{3}N_{4}}$ on a clean thermal wet $\mathrm{SiO{{}_2}}$
substrate using low-pressure chemical vapor deposition (LPCVD). Deep-ultraviolet
(DUV) stepper lithography (ASML PAS850C) with $110\,\mathrm{nm}$
resolution was used to patten the circuits, followed by inductively
coupled plasma (ICP) etching and thermal annealing. Afterwards, $\mathrm{Si_{3}N_{4}}$
annealing at $1200^{\mathrm{o}}\mathrm{C}$ was performed to remove
hydrogen content in $\mathrm{Si_{3}N_{4}}$, followed by 3-$\mathrm{\mu m}$-thick
$\mathrm{SiO{{}_2}}$ top cladding deposition and another thermal
annealing to eliminate hydrogen contents in $\mathrm{SiO{{}_2}}$.
Finally, UV photolithography and deep dry etching were performed to
define chip size and create smooth chip facets. The wafer was separated
into individual chips for experiments.

\section{experimental results}

Figure \ref{fig:2}A illustrates the experimental setup, with the
LN modulator chip and $\mathrm{Si_{3}N_{4}}$ microring resonator
chip shown in Fig. \ref{fig:2}B. To achieve maximum compression
of the pulsed light, the LN chip and microring resonator chip are
connected by a 400-meter-long single-mode fiber. Based on the dimensions
of the $\mathrm{Si_{3}N_{4}}$ microring resonator, the modulation
signal applied to the phase modulator is set to approximately $35.5\,\mathrm{GHz}$
to match the free spectral range (FSR) of the microring resonator.
The spectrum of the EO comb generated after the CW laser passes through
the LN chip is shown in Fig. \ref{fig:2}C. It can be observed that
the center frequency of the EO comb shifts with changes in the CW
pump frequency, while the spectral width remains stable at approximately
$4.6\,\mathrm{nm}$. The corresponding temporal pulse profile, which
is obtained via autocorrelation measurements, is shown in Fig. \ref{fig:2}D,
with a pulse width of approximately $2.5\,\mathrm{ps}$. The pulse
is coupled into the microring resonator through the fiber lens with
a coupling loss of $3.0\,\mathrm{dB}$ per facet. Figure \ref{fig:2}E
presents the typical resonance mode of the microring resonator, where
the fitted loaded Q-factor is approximately $8.4\times10^{6}$. Additionally,
the integrated dispersion curve of the microring resonator, which
is measured using a Mach-Zehnder interferometer, is shown in Fig.
\ref{fig:2}F. The results reveal anomalous dispersion near 1560 nm,
with $D_{2}/2\pi\approx3\,\mathrm{MHz}$.

\begin{figure*}[t]
\centering{}\includegraphics[clip,width=2\columnwidth]{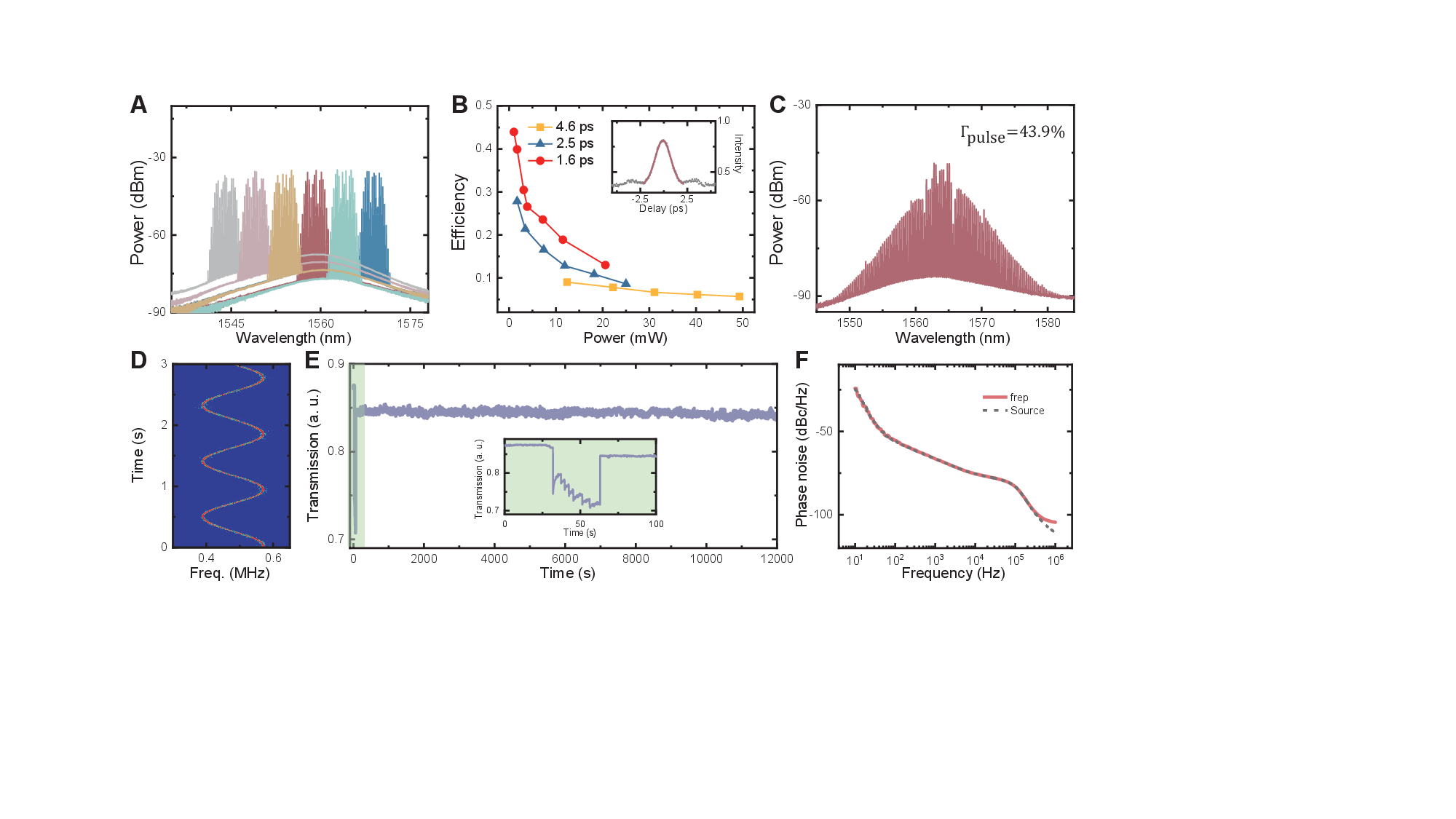}\caption{\label{fig:4} \textbf{(A)} Optical spectrum of the cascaded EO comb.
\textbf{(B)} The convension efficiency of the DKSs with various pump
power and different pulse width. Inset: the autocorrelation measurement
of the cascaded EO comb in (A).\textbf{ (C)} Optical spectrum of the
DKS with the $43.9
$ convension efficiency. \textbf{(D)} The real-time evolution of the
repetition rate while scanning the modulation frequency.\textbf{ (E)}
Time-series measurements of transmission of the pump laser. \textbf{Inset:
}Enlarge of the time-series measurement at $0-100\,\mathrm{s}$. \textbf{(F)}
Phase noise measurement of the repetition rate and the RF source.
The phase noise of the RF source and the repetition rate are at the
same level.}
\end{figure*}

The generation of soliton microcombs requires the pump laser to be
stabilized on the soliton step on the red-detuned side of the resonance
mode. However, the traditional CW pumping scheme is hindered by intracavity
thermal effects, making it challenging to achieve a stable soliton
state via simple wavelength scanning. Figure \ref{fig:2}G depicts
the intracavity power spectrum (purple curve) when the CW pump laser
is scanned from the blue-detuned side to the red-detuned side. Only
a minimal soliton step can be observed. In contrast, with pulse pumping,
the improved conversion efficiency results in a pronounced soliton
step in the intracavity power spectrum (red curve). This allows for
stable soliton states to be achieved by simply scanning the pump laser
from the blue-detuned to the red-detuned side. Figure \ref{fig:2}H
compares the soliton spectra generated by pulse pumping and CW pumping
(using a dual-pumped method). Due to the significantly enhanced conversion
efficiency, the soliton state generated by pulse pumping exhibits
higher comb line power and broader spectral width.

In the case of pulse pumping, precise matching between the modulation
frequency of the phase modulator (i.e., the repetition rate of the
EO comb) and the repetition rate of the soliton microcomb is critical.
Figure \ref{fig:3}A illustrates the changes in the pump laser transmission
spectrum as the modulation frequency increases from $35.535467\,\mathrm{GHz}$
to $35.536467\,\mathrm{GHz}$. As the modulation frequency increases,
the matching between the modulation frequency and the soliton microcomb
repetition rate improves, resulting in a broader soliton step. When
the modulation frequency reaches $35.535867\,\mathrm{GHz}$, the best
matching is achieved, and the soliton step reaches its maximum width.
Furthermore, the optimal modulation frequency is influenced by the
pump power due to intracavity thermal effects. As shown in Fig. \ref{fig:3}B,
an increase in pump power leads to a reduction in the FSR of the microring
resonator caused by thermal effects, which lowers the repetition rate
of the soliton microcomb and, consequently, the optimal modulation
frequency. Additionally, the pump laser power affects the conversion
efficiency, as illustrated in Fig. \ref{fig:3}C. The conversion efficiency
decreases with increasing pump laser power. According to Eq. (\ref{eq:eq2}),
the conversion efficiency is primarily determined by the combline
number of the EO comb. While an increase in optical power does not
affect these comb lines, it instead raises the total pump power. Consequently,
the conversion efficiency decreases as the optical power increases
\cite{li2024breaking}. When the pulse width is $2.5\,\mathrm{ps}$,
the conversion efficiency can reach $27.8\text{\,\%}$. Here, we adopt
the widely accepted definition of \textquotedbl conversion efficiency,\textquotedbl{}
which is the ratio of the power within the comb lines (excluding the
power carried by the pump frequencies) to the pump laser power on
chip \cite{yang2024efficient}. More details of the estimation of
conversion efficiency see Supplementary Information. 

\begin{table*}
\begin{tabular*}{2\columnwidth}{@{\extracolsep{\fill}}>{\centering}m{2cm}>{\centering}m{3cm}>{\centering}m{3cm}>{\centering}m{2.7cm}>{\centering}m{2.7cm}}
\hline 
Comb generation platform & Repetition rate (GHz) & Pump scheme & Conversion efficiency & Soliton state\tabularnewline
\hline 
\hline 
SiO$_{2}$ \cite{li2022efficiency} & 22.1 & fiber EO comb & 34\% & yes\tabularnewline
\hline 
$\mathrm{Si_{3}N_{4}}$ \cite{anderson2021photonic} & 27.9 & fiber EO comb & 8\% & yes\tabularnewline
\hline 
LN\cite{cheng2024frequency} & 30.1 & on chip pulse & 10\% & no\tabularnewline
\hline 
our work & 35.5 & on chip pulse & 43.9\% & yes\tabularnewline
\hline 
\end{tabular*}

\caption{\label{table:1} Performance with other reported pulse pump high conversion
efficiency Kerr combs.}
\end{table*}

To further enhance the conversion efficiency, we cascaded two on-chip
phase modulators to broaden the spectral width of the EO comb. After
cascading, the spectral width of the EO comb is extended to $6.2\,\mathrm{nm}$,
with a significantly flatter spectral profile, and the corresponding
pulse width is reduced to $1.6\,\mathrm{ps}$, as shown in Fig. \ref{fig:4}A.
Figure \ref{fig:4}B illustrates the variation in conversion efficiency
with pump power under different pulse widths. It can be observed that,
at the same pump power, narrower pulse widths result in higher conversion
efficiencies, consistent with the prediction of Eq. (\ref{eq:eq1}).
Additionally, narrower pulses provide higher peak power for the same
pump power, leading to lower soliton microcomb generation thresholds.
When the pulse width is reduced to $1.6\,\mathrm{ps}$, a pump power
as low as $1\,\mathrm{mW}$ is sufficient to generate a soliton microcomb,
achieving a conversion efficiency of $43.9\,\%$. The corresponding
soliton spectrum is shown in Fig. \ref{fig:4}C. This power level
can be easily achieved with a distributed feedback (DFB) laser, making
it feasible to construct fully integrated soliton microcomb with high
conversion efficiency.

For quantitative characterization of repetition rate tunability, we
implemented continuous sinusoidal modulation on the RF driving signal
while simultaneously tracking the repetition rate shifts. The real-time
measurement in Fig. \ref{fig:4}D demonstrates a dynamic modulation
range exceeding 200 kHz peak-to-peak. To fully characterize the high-conversion-efficiency
DKS, it is essential to evaluate its long-term stability. We recorded
the frequency variation of the pump transmission using a data acquisition
card with a gate time of $0.1\,\mathrm{s}$. Figure \ref{fig:4}E
shows the pump laser transmission over a period of $12000\,\mathrm{s}$.
At the beginning of the time series (shaded area), the pump laser
frequency is manually tuned from the blue side to the red side of
the cavity resonance. The enlarged time series is shown in inset of
Fig. \ref{fig:4}E. Furthermore, we also measured phase noise of the
repetition rate and the RF signal using a phase noise analyzer (Rohde
\& Schwarz, FSWP). The results are shown in Fig. \ref{fig:4}F, confirming
that the phase noise of the RF signal and the repetition rate are
at the same level.

\section{Discussion}

In conclusion, we have demonstrated a photonic chip-based high-efficiency
soliton microcomb system that synergistically combines the exceptional
EO modulation efficiency of LN with the ultralow-loss and dispersion-engineered
microresonators of $\mathrm{Si_{3}N_{4}}$. By modulating the CW pump
source with the on-chip LN phase modulator and compressing the resulting
pulses through dispersion-compensating fiber, we generate $1.6\,\mathrm{ps}$
optical pulses for pumping soliton microcmbs. This pulse-pumping scheme
achieves $43.9\,\%$ conversion efficiency at an average on-chip pump
power of only $1\,\mathrm{mW}$, surpassing all prior pulsed microcomb
systems in efficiency (Table \ref{table:1}). Crucially, the approach
reduces the soliton threshold power, suppresses thermal instabilities,
and enables robust soliton generation through straightforward pump
frequency scanning. These advances resolve key challenges in power
efficiency and operational complexity that have hindered practical
microcomb applications.

While our proof-of-concept currently employs discrete components,
recent breakthroughs in heterogeneous and hybrid photonic integration
provide a clear roadmap toward full system integration. The sub-mW
operational threshold enables direct driving by heterogeneous III-V
DFB lasers \cite{shen2020integrated}, bypassing external amplifiers,
as demonstrated in recent progress in laser integration on $\mathrm{Si_{3}N_{4}}$
\cite{xiang2021laser} and LN \cite{li2022integrated} platforms.
Moreover, dispersion engineering of $\mathrm{Si_{3}N_{4}}$ or LN
waveguides allows on-chip pulse compression, eliminating reliance
on bulk fiber components \cite{yu2022integrated}. The rapid maturation
of multi-material integration techniques facilitates the monolithic
integration of LN modulators, $\mathrm{Si_{3}N_{4}}$ dispersion compensators,
and III-V pump lasers on a unified platform through III-V/ $\mathrm{Si_{3}N_{4}}$/LN
heterogeneous bonding \cite{churaev2023heterogeneously}. Such integration,
leveraging fast modulation of LN and low-loss optics of $\mathrm{Si_{3}N_{4}}$,
establishes a scalable platform for deployable microcombs in portable
sensors, optical clocks, and communication systems.

\vbox{}

\noindent\textbf{Data availability}\\ All data needed to evaluate
the conclusions in the paper are present in the paper and/or the Supplementary
Information.

\vbox{}

\noindent\textbf{Acknowledgments}\\ The work was supported by the
the National Natural Science Foundation of China (12293050, 12293052,
12104442, 12304435, 12261131503, 92050109, 62225506 and 92250302),
Innovation program for Quantum Science and Technology (2021ZD0303203,
2023ZD0301500), the CAS Project for Young Scientists in Basic Research
(YSBR-069), the Postdoctoral Fellowship Program of CPSF (GZC20232564),
the China Postdoctoral Science Foundation (2023M733414), Shenzhen
Science and Technology Program (Grant No. RCJC20231211090042078),
the Fundamental Research Funds for the Central Universities. This
work was partially carried out at the USTC Center for Micro and Nanoscale
Research and Fabrication.

\vbox{}

\noindent\textbf{\textcolor{black}{Author contributions}}\textcolor{black}{\\
R.N. and S.W. contribute equally to this work. R.N., S.W. and C.-H.D.
conceived the experiments. S.W., P.-Y.W., R.M., J.L. and F.B. prepared
devices. R.N. built the experimental setup and carried out measurements,
with the assistance from F.-W.S.. R.N. and S.W. analyzed the data,
with the assistance from J.L. and Z.S.. R.N., S.W., and C.-H.D. wrote
the paper with input from all co-authors. C.-H.D. and G.-C.G. supervised
the project. All authors contributed extensively to the work presented
in this paper.}

\textcolor{brown}{\emph{}}

\vbox{}

\noindent\textbf{Competing financial interests}\\The authors declare
no competing financial interests.

\bibliographystyle{Microcavity}
\phantomsection\addcontentsline{toc}{section}{\refname}\bibliography{Manuscript}

\end{document}